\documentclass[french]{pfia}
\usepackage{cite}
\usepackage{amsmath,amssymb,amsfonts}
\newtheorem{definition}{Definition}
\usepackage{algorithmic}
\usepackage{graphicx}
\usepackage{textcomp}
\usepackage{xcolor}
\usepackage{seqsplit}
\usepackage{tikz}
\usepackage{tabularx}
\usepackage{array}
\usepackage{listings}
\usepackage{adjustbox}
\usepackage{multirow}
\usepackage[shortlabels]{enumitem}
\usepackage{comment}
\usepackage{titlesec}
\titlespacing*{\section}{0pt}{0.1\baselineskip}{0.1\baselineskip}

\newcolumntype{M}[1]{>{\centering\arraybackslash}m{#1}}
\usetikzlibrary{shapes.geometric,shapes.symbols,fit,positioning,shadows, arrows.meta, backgrounds}

\def\BibTeX{{\rm B\kern-.05em{\sc i\kern-.025em b}\kern-.08em
		T\kern-.1667em\lower.7ex\hbox{E}\kern-.125emX}}

\title{\textbf{Construction d'un système de recommandation basé sur des contraintes via des graphes de connaissances}}

\author{Ngoc Luyen Le\fup{1,2}, Marie-Hélène Abel\fup{1}, Philippe Gouspillou\fup{2}\\[6pt]
\fup{1} Université de technologie de Compiègne, CNRS, Heudiasyc\\(Heuristics and Diagnosis of Complex Systems), CS 60319 - 60203 Compiègne Cedex, France.\\
\fup{2} Vivocaz, 8 B Rue de la Gare, 02200, Mercin-et-Vaux, France.}

\date{}

\begin{document}

\maketitle


\begin{resume}
Les graphes de connaissances en RDF modélisent des entités et leurs relations via des ontologies. Leur utilisation a gagné en popularité pour la modélisation de l'information, notamment dans les systèmes de recommandation où les éléments d'information et les utilisateurs sont intégrés dans ces graphes, révélant davantage de liens et de relations. Les systèmes de recommandation basés sur des contraintes exploitent une connaissance approfondie des recommandations pour identifier celles pertinentes. En les combinant avec des graphes de connaissances, nous obtenons plusieurs avantages en termes d'ensembles de contraintes. Cet article explore la construction d'un système de recommandation basé sur des contraintes avec des graphes de connaissances RDF dans le domaine de l'achat/vente de véhicules. Nos expérimentations démontrent que l'approche proposée identifie efficacement des recommandations selon les préférences des utilisateurs.
\end{resume}

\begin{motscles}
Graphe de connaissances, Système de recommandation basé sur des contraintes, Ontologie
\end{motscles}

\begin{abstract}
Knowledge graphs in RDF model entities and their relations using ontologies, and have gained popularity for information modeling. In recommender systems, knowledge graphs help represent more links and relationships between users and items. Constraint-based recommender systems leverage deep recommendation knowledge to identify relevant suggestions. When combined with knowledge graphs, they offer benefits in constraint sets.
This paper explores a constraint-based recommender system using RDF knowledge graphs for the vehicle purchase/sale domain. Our experiments demonstrate that the proposed approach efficiently identifies recommendations based on user preferences.
\end{abstract}



\section{Introduction}
Dans plusieurs domaines tels que les services financiers, les biens de luxe, l'immobilier ou les automobiles, les achats généralement plus coûteux sont moins fréquents  que les achats de commodité. Par conséquent, faire des recommandations de ce type de produit nécessite l'obtention de plus d'informations provenant des utilisateurs tels que leurs préférences ou leurs besoins. En d'autres termes, le système de recommandation (SdR) tente de récupérer des éléments pertinents, provenant des réponses aux questions sur ses besoins et préférences, pour établir les recommandations les plus appropriées. Par conséquent, les SdRs à base de contraintes présentent une approche typique pour répondre à ce type de domaine applicatif.

Dans les SdRs à base de contraintes, l'identification des recommandations est considérée comme un processus de satisfaction des contraintes. Certaines contraintes peuvent provenir de la définition et la connaissance du domaine concerné par l'élément, l'item à considérer pour le recommandation. D'autres contraintes peuvent se définir à partir du profil de l'utilisateur comme ses préférences \cite{felfernig2008constraint}. La combinaison de deux types de contraintes provoque une augmentation de l'espace de recherche d'un élément. Par ailleurs, cela peut conduire à la répétition du même type de contraintes pour des groupes d'utilisateurs partageant certaines caractéristiques communes. L'utilisation de graphes de connaissances RDF avec le support d'ontologies peut aider à réduire l'ensemble de contraintes en utilisant des mécanismes de raisonnement pour déduire des informations pertinentes sur les connaissances spécifiques à un domaine. Dans cet article, nous présentons notre approche pour la construction d'un système de recommandation basé sur des contraintes via des graphes de connaissances.

Le reste de cet article est organisé comme suit. Dans la section suivante, nous présentons des travaux de la littérature sur les SdRs basés sur des contraintes. La section \ref{notr_proposition} présente nos principales contributions à la construction du SdR basé sur des contraintes exploitant les graphes de connaissances RDF. Dans la section \ref{experiments}, une expérimentation sur le domaine de l'achat/vente de véhicules de notre approche est présentée et discutée. Pour finir, nous concluons et avançons quelques perspectives. 

\section{Travaux de la littérature}\label{sec_travaux_litterature}
Les SdRs sont une application spéciale qui estime la préférence des utilisateurs pour les éléments/items et tente de recommander les éléments les plus pertinents aux utilisateurs via la récupération d'informations\cite{lu2015recommender}. Les recommandations effectuées visent à aider les utilisateurs dans divers processus de prise de décision tels que la musique à écouter ou les produits à acheter. En général, les SdRs sont généralement classés en six catégories principales : les SdRs basés sur le filtrage collaboratif, les SdRs basés sur le contenu, les SdRs basés sur la démographie, les SdRs basés sur les connaissances, les SdRs sensibles au contexte et les SdRs hybrides \cite{le2022towards}.

Si la quantité de données collectées est limitée, les résultats des systèmes tels que les SdRs de filtrage collaboratif, les SdRs basés sur le contenu et les SdRs démographiques peuvent être pauvres ou ne pas couvrir complètement le spectre des combinaisons entre les utilisateurs et les éléments. En effet, ces approches peuvent rencontrer des problèmes tels que le démarrage à froid, la rareté des données, l'analyse limitée du contexte et la spécialisation excessive \cite{adomavicius2005toward, ramezani2008selecting}. Les SdRs basés sur la connaissance sont proposés pour résoudre ces problèmes en sollicitant explicitement les préférences des utilisateurs pour de tels éléments et en utilisant une connaissance approfondie du domaine pour calculer des recommandations pertinentes \cite{felfernig2011developing}. En particulier, ce type de SdR convient bien aux situations où (i) les utilisateurs souhaitent spécifier explicitement leurs exigences ; (ii) il est difficile d'obtenir des commentaires sur les éléments ; et (iii) les commentaires peuvent être obsolètes ou sensibles au temps. Par exemple, si un élément est une voiture d'occasion, les commentaires peuvent ne pas être très utiles pour calculer des recommandations car une voiture d'occasion n'est achetée qu'une seule fois.

En considérant la manière dont les utilisateurs interagissent et la base de connaissances correspondante utilisée pour ces interactions, il existe deux types de SdRs basés sur la connaissance : les SdRs basés sur les contraintes \cite{felfernig2015constraint} et les SdRs basés sur les cas \cite{bridge2005case}. Alors que les SdRs basés sur les cas trouvent des éléments similaires en calculant et en adaptant les recommandations en fonction des cas similaires dans le passé, les SdRs basés sur les contraintes définissent un ensemble de règles/contraintes pour faire correspondre les préférences/les exigences des utilisateurs aux propriétés des éléments. Les SdRs basés sur les contraintes ont été appliqués dans différents domaines pour aider les utilisateurs à adopter les meilleures recommandations d'éléments pertinents. Dans \cite{boudaa2021datatourist, jannach2009constraint}, les auteurs ont développé des SdRs basés sur les contraintes en se basant sur l'utilisation de bases de connaissances dans le domaine du tourisme. Dans \cite{atas2019towards}, l'auteur a proposé une amélioration de l'utilisation des SdRs basés sur les contraintes en utilisant la similarité des exigences des utilisateurs. L'utilisation de règles/contraintes est devenue de plus en plus populaire pour améliorer les résultats des recommandations, comme dans les applications e-commerce \cite{dadouchi2022context}, les systèmes de simulation \cite{le2023constraint} ou les services financiers \cite{felfernig2016application}.

L'achat et la vente de véhicules d'occasion n'est pas aussi fréquent que d'autres produits, et chaque véhicule n'a qu'une seule transaction. En général, les préférences des utilisateurs pour leurs véhicules préférés jouent un rôle important dans la recommandation de véhicules d'occasion pertinents. Afin d'effectuer les recommandations les plus pertinentes pour ce type de transaction, nous avons choisi de construire un SdR à base de contraintes en s'appuyant sur des graphes de connaissances. Dans la section suivante, nous présenterons en détail notre approche pour ce travail.
\section{Notre approche} \label{notr_proposition}
Dans cette section, nous présentons notre approche pour la construction d'un SdR à base de contraintes s'appuyant sur un graphe de connaissances RDF. Pour illustrer notre approche, nous utilisons des ontologies du domaine de l'e-commerce liées à l'achat et à la vente de véhicules pour créer une base de connaissances.
\subsection{Graphe de connaissances via RDF}
La construction d'une base de connaissances pour le domaine des véhicules se compose de trois axes principaux : les propriétés des véhicules, les profils des utilisateurs-acheteurs et les interactions entre les utilisateurs-acheteurs et les véhicules.
La collecte de ces informations peut être organisée et réécrite sous forme de triplets, définis formellement comme $G_V$ = $\{a_{1}^{v},$ $a_{2}^{v},$ $...,$ $a_{n}^{v}$$\}$ où $a_{i}^{v}$ représente un triplet RDF complet $a_{i}^{v} = \langle sujet_i, prédicat_i, objet_i\rangle$. De même, les profils d'utilisateurs qui comprennent des informations sur les utilisateurs et leurs préférences en matière de véhicules peuvent également être définis comme un ensemble de triplets RDF : $G_U$ $=$ $\{a_{1}^{u},$ $a_{2}^{u},$ $...,$ $a_{m}^{u}\}$ où $a_{j}^{v}$ représente un triplet RDF complet. Enfin, lorsqu'un utilisateur ajoute un élément à sa liste d'éléments préférés, cela signifie que cet élément est intéressant pour l'utilisateur. Ces interactions entre les utilisateurs et les éléments sont définies comme : $RS$ $:$ $G_U$ $\times$ $G_V$ $\times$ $G_C$ $\longrightarrow$ $Interaction$ où $G_U$ correspond à l'utilisateur, $G_V$ désigne la description du véhicule et $G_C$ exprime des informations contextuelles concernant l'utilisateur et l'élément lorsque l'interaction est effectuée, par exemple, les objectifs de l'utilisateur, la date, le lieu et les informations sur les ressources. Dans notre travail, nous utilisons l'ontologie développée pour la description des véhicules et des profils utilisateurs présentée dans \cite{le2022towards}.

\subsection{Système de recommandation basé sur les contraintes}
Après avoir défini les bases d'un graphe de connaissances RDF pour l'achat/la vente de véhicules, nous montrons dans cette section comment définir et construire un SdR basé sur des contraintes à partir de cette source de données.

 Dans notre travail, nous nous concentrons sur le traitement des exigences des utilisateurs à partir de leurs préférences et de leurs informations contextuelles. Tout d'abord, les préférences des utilisateurs concernant leur véhicule préféré sont considérées comme une partie des informations dans les profils d'utilisateurs. Par conséquent, les utilisateurs doivent fournir leurs préférences relatives aux caractéristiques du véhicule qu'ils aimeraient posséder. Par exemple, plusieurs utilisateurs peuvent avoir une préférence pour la couleur \textit{noire} ou \textit{blanche} pour leur véhicule, ou d'autres utilisateurs veulent un \textit{véhicule avec 7 places pour la famille}. Deuxièmement, les informations contextuelles de l'utilisateur peuvent être les situations externes. Par exemple, l'endroit où les utilisateurs vivent ou travaillent peut être un facteur important dans la sélection des types de véhicules. Par conséquent, les informations sur les préférences des utilisateurs et le contexte de l'utilisateur jouent un rôle de contraintes afin de filtrer les éléments de recommandation pertinents pour les utilisateurs.

D'autre part, nous établissons le pont entre les exigences des utilisateurs et les éléments de description des véhicules en utilisant les descriptions de véhicules et la connaissance du domaine. Tout d'abord, la description du véhicule englobe les propriétés d'un élément donné, tandis que la connaissance du domaine fournit des informations plus approfondies sur les éléments. Par exemple, lorsqu'un utilisateur déclare son profil et exprime son intérêt pour un ``profil de famille'', la connaissance du domaine pour les éléments de véhicule permet la recommandation de véhicules de grande taille qui ont \textit{un nombre de places} supérieur à \textit{trois sièges}.

La recommandation basée sur les contraintes repose sur l'exploration des relations entre les exigences de l'utilisateur et les propriétés de l'élément. La base de connaissances dans notre cas peut être considérée comme un ensemble de variables et un ensemble de contraintes. L'utilisation de ces variables et contraintes peut constituer les éléments d'un Problème de Satisfaction de Contraintes (CSP) \cite{felfernig2006integrated, felfernig2011developing}. Les solutions de ce CSP permettent de trouver les recommandations les plus pertinentes dans un SdR. La tâche de calcul et de suggestion des recommandations pour un utilisateur en fonction de ses préférences est appelée une tâche de recommandation. 

\begin{definition} La tâche de recommandation est définie comme un CSP($\mathcal{V}_U$,$\mathcal{V}_I$,$\mathcal{C}$), où $\mathcal{V}_U$ =$\{vu_1, vu_2, ..., vu_n\}$ désigne un ensemble de variables qui représentent les préférences de l'utilisateur, $\mathcal{V}_I$ = $\{vi_1, vi_2, ..., vi_m\}$ est un ensemble de variables qui représentent les propriétés des éléments, $\mathcal{C} = \mathcal{C}_{KB} \cup \mathcal{C}_F$ fait référence à l'ensemble des contraintes représentant les contraintes spécifiques au domaine $\mathcal{ C}_{KB}$ et l'ensemble de contraintes de filtre $\mathcal{C}_F$ qui décrivent le lien entre les préférences de l'utilisateur et les éléments.
	\label{def1}
\end{definition}

Dans le cadre d'une application e-commerce d'achats/ventes de véhicules, nous pouvons extraire différentes préférences utilisateur sous la forme d'un ensemble de variables pour $\mathcal{V}_U$ et les propriétés des éléments du véhicule sous la forme de l'ensemble de variables pour $\mathcal{V}_I$. En particulier, nous illustrons les ensembles de variables par un exemple simple comme suit :

\begin{itemize}[\textbf{+}]
	\item  $\mathcal{V}_U$ = $\{ vu_1: typeDeV\acute{e}hicule(sedan,suv,van)$,\newline $vu_2:couleur(bleu, noir, blanc, rouge)$,\newline
	$vu_3:profil(utilisateurEtudiant,\newline utilisateurParent, profilProfessionnel)$,\newline
	$vu_4:nombreDeSi\grave{e}ges(entier)$,\newline  $vu_5:maxKilom\acute{e}trage(entier)$, 
	\newline $vu_6:marque(texte)$,
	 $vu_{7}:maxBudget(entier)\}$
	\item  $\mathcal{V}_I$ = $\{ vi_1: nom (texte)$, $vi_2:prix(entier)$,\newline $vi_3:typeDeCarrosserie(texte)$,
	\newline $vi_4:nombreDeSi\grave{e}ges(entier)$, 	\newline $vi_5:ann\acute{e}eDuMod\grave{e}le(2021,2020, 2019,2018)$, 
	\newline $vi_6:marque(Peugeot, Renault, Citroen)$, 
	\newline $vi_7:kilom\acute{e}trage(entier)\}$
\end{itemize}

Chaque contrainte peut être classée en $\mathcal{C}_{KB}$ ou $\mathcal{C}_F$. Alors que les contraintes $\mathcal{C}_{KB}$ sont formées à partir de la connaissance du domaine, $\mathcal{C}_F$ définit les exigences particulières de l'utilisateur sur les éléments. Nous montrons plusieurs exemples de contraintes $\mathcal{C}_{KB}$ et $\mathcal{C}_F$ dans le tableau  \ref{table_02}.

\begin{table}[h]
	\begin{center}
		\begin{tabular}{|p{0.7cm} | p{6.5cm} |}
			\hline 
			\textbf{ID} & \textbf{Description de la contrainte} \\\hline
			$\mathcal{C}_{KB1}$ & Une inspection technique datant de moins de 6 mois est requise pour un véhicule d'occasion de plus de 4 ans.    \\\hline
			$\mathcal{C}_{KB2}$ & Si les utilisateurs préfèrent les longs trajets, un SUV ou un Crossover peut leur convenir.    \\\hline
			$\mathcal{C}_{F1}$ & le prix de véhicule doit être inférieur ou égal au budget maximal de l'utilisateur.  \\\hline
			$\mathcal{C}_{F2}$ & le nombre de kilomètres parcourus par le véhicule doit être inférieur au kilométrage maximal imposé par l'utilisateur.   \\\hline
			$\mathcal{C}_{F3}$ & le nombre de places du véhicule doit être égal au nombre de sièges requis par l'utilisateur. \\\hline
			$\mathcal{C}_{F4}$ & la couleur du véhicule doit être soit blanche soit bleue.  \\\hline
		\end{tabular}
	\end{center}
	\vspace{-0.5cm}
	\caption{Exemple de contraintes liées aux connaissances spécifiques au domaine et aux préférences de l'utilisateur}
	\vspace{-0.5cm}
	\label{table_02}
\end{table}

\begin{definition} Une recommandation (une solution) pour une tâche de recommandation donnée ($\mathcal{V}_U$,$\mathcal{V}_I$,$\mathcal{C}$) est définie comme une instanciation de $ \mathcal{V}_I$ en réalisant une affectation complète aux variables de ($\mathcal{V}_U$,$\mathcal{V}_I$) telle que les contraintes en $\mathcal{C}$ soient satisfaites. La recommandation est \textit{cohérente} si les affectations sont \textit{cohérentes} avec les contraintes.
\label{def2}
\end{definition}
Les SdRs basés sur des contraintes reposent sur une base de connaissances explicite du domaine des utilisateurs et des éléments. Avec deux types de contraintes, nous pouvons calculer des recommandations pertinentes pour un utilisateur. Les contraintes de $\mathcal{C}_{KB}$ liées à la connaissance spécifique du domaine peuvent être satisfaites en utilisant des règles qui sont intégrées dans des ontologies. Par conséquent, nous explorons cette approche dans la prochaine section en se basant sur le modèle d'ontologie repris de \cite{le2022towards, le2022apport} et le graphe de connaissances RDF pour les profils d'utilisateurs et les descriptions de véhicules.
\subsection{Contraintes de connaissance spécifiques au domaine par des règles SWRL}
Dans le contexte du domaine de l'achat/vente de véhicules, des ontologies sont utilisées pour structurer et organiser les descriptions de véhicules et les profils d'utilisateurs. L'ontologie proposée est construite à l'aide du langage d'ontologie Web (OWL) \cite{mcguinness2004owl, le2022towards}, qui est un langage de représentation des connaissances hautement expressif, flexible et efficace basé sur l'arrière-plan mathématique de la logique de description. OWL peut réaliser le raisonnement sur les informations implicites en traitant des connaissances explicites, ce qui améliore la gestion de l'information. 
Les règles sont utiles pour implémenter la partie déductive de la base de connaissances. Dans ce travail, nous utilisons le langage de règles du web sémantique  (SWRL) pour écrire des règles sur des graphes de connaissances RDF.

Les contraintes dans l'ensemble de contraintes $\mathcal{C}_{KB}$ s'appliquent souvent à une classe, aux propriétés d'une classe ou à un groupe d'individus. En d'autres termes, ces contraintes affectent les informations globales dans le cadre de la base de connaissances. Ces contraintes peuvent être traduites en règles à intégrer dans l'ontologie à l'aide de SWRL. Par exemple, pour la contrainte $\mathcal{C}_{KB2}$ qui est utilisée pour tous les utilisateurs ayant une préférence pour la \textit{route longue distance}, nous pouvons utiliser une règle SWRL pour déduire le \textit{type de véhicule} préféré par l'utilisateur. Par conséquent, nous proposons de représenter les contraintes de connaissances spécifiques au domaine à l'aide de règles SWRL, en se basant sur les avantages en matière de déduction d'informations.

\begin{table}[h]
	\begin{center}
		\vspace{-0.2cm}
		\begin{adjustbox}{width=0.46\textwidth,center}
			\begin{tabular}{|m{2.8em} | m{9cm}|}
				\hline
				\textbf{ID} & \textbf{Expression de règle SWRL} \\\hline
				$\mathcal{C}_{KB1}$ &  
				$Automobile(?a)$ $\wedge$ $Contr\hat{o}leTechnique(?c)$ $\wedge$ $inspect\acute{e}(?a,$ $?c)$   $\wedge$ $DateDeProduction(?a,$ $?pdate)$ $\wedge$ $valideDe(?c,$ $?cdate)$ $\wedge$ $temporal:duration$$(?pdur\acute{e}e,$ $?pdate,$ $``maintenant",$ $``mois")$ $\wedge$ $temporal:duration$$(?cdur\acute{e}e,$ $?cdate,$ $``maintenant",$ $``mois")$ $\wedge$ $swrlb:greaterThan$$(?pdur\acute{e}e,$ $48)$ $\wedge$ $swrlb:greaterThan$$(?cdur\acute{e}e,$ $6)$ $\longrightarrow$ $estRequis$$(?c,$ $vrai)$
				\\\hline
				$\mathcal{C}_{KB2}$ &  $Pr\acute{e}f\acute{e}renceDeV\acute{e}hicule(?vpu)$ $\wedge$ $aLeTypeDeRoutePr\acute{e}f\acute{e}r\acute{e}(?vpu,$ $?route)$ $\wedge$ $sameAs(?route,$ $upo:longDistanceRoute)$ $\longrightarrow$ $aUnTypeDeV\acute{e}hiculePr\acute{e}f\acute{e}r\acute{e}(?vpu,$ $upo:SUV)$ $\wedge$ $aUnTypeDeV\acute{e}hiculePr\acute{e}f\acute{e}r\acute{e}(?vpu,$ $upo:Crossover)$\\
				\hline
			\end{tabular}
		\end{adjustbox}
	\end{center}
	\vspace{-0.5cm}
	\caption{Les règles SWRL pour les contraintes définies dans le tableau \ref{table_02}.}
	\label{tab:rule_tb1}
	\vspace{-0.3cm}
\end{table}

Les règles SWRL offrent de puissantes capacités déductives exploitant une modélisation ontologique. Cependant, SWRL est essentiellement un langage de règles, et il ne fournit pas de support solide pour filtrer et interroger les informations du graphe de connaissances RDF. Par conséquent, nous présenterons une approche pour les contraintes $\mathcal{C}_{F}$ liées aux préférences de l'utilisateur qui implique le filtrage et la mise en correspondance sur les graphes de connaissances RDF dans la section suivante.

\subsection{Contraintes de préférence de l'utilisateur par des requêtes SPARQL}

Supposons que $Q$ est une requête SPARQL et que $c$ est une contrainte. $Q$ FILTER $c$ est appelée une requête de contrainte, où chaque variable dans la contrainte est satisfaite dans la requête $Q$. Une solution d'une requête SPARQL $Q$ est définie comme une assignation de variables dans $Q$ à des valeurs. Un ensemble de valeurs possibles qui peuvent être assignées à une variable est appelé un domaine. Une recommandation ou solution est cohérente si toutes les variables déclarées dans la requête ont une valeur correspondante garantie. Pour trouver toutes les solutions possibles, nous sélectionnons une valeur dans le graphe de connaissances RDF pour chaque variable et nous assurons qu'elle vérifie les conditions des motifs et des filtres. En tenant compte de ces aspects, trouver des recommandations pour le SdR basé sur des contraintes défini dans les définitions \ref{def1} et \ref{def2} revient à trouver les solutions d'une requête SPARQL $Q$ avec un ensemble de contraintes $c$. Les expressions équivalentes de celles-ci sont décrites :

\begin{itemize}[\textbf{+}]
	\item Les variables dans $\mathcal{V}_U$ et $\mathcal{V}_I$ sont utilisées comme variables principales dans la requête SPARQL $Q$ sur les graphes de connaissances RDF associés à $G_U$ et $G_I$.
	\item Les contraintes $c \in \mathcal{C}_F$ doivent être satisfaites en incorporant la clause FILTER dans la requête SPARQL $Q$. 
\end{itemize} 
\begin{figure}[h]
	\vspace{-0.4cm}
\begin{adjustbox}{width=.39\textwidth,center}
	\begin{lstlisting}[language=SPARQL, label=lst:sparql,basicstyle=\scriptsize\ttfamily,frame=lines, numbers=left,firstnumber=1,xleftmargin=2.5em,framexleftmargin=2.5em,mathescape=true]
 PREFIX uvso: <http: //utc.fr/uvso/ns#>
 PREFIX uvo: <http: //utc.fr/uvo/ns#> 
 PREFIX uvoo: <http: //utc.fr/uvoo/ns#>
 PREFIX rdf: <http: //w3.org/1999/02/22-rdf-syntax-ns#> 
 PREFIX xsd: <http: //w3.org/2001/XMLSchema#> 
 PREFIX gr: <http: //purl.org/goodrelations/v1#> 
			
 SELECT ?auto
 WHERE {
	?auto rdf:type uvso:Automobile.
	?auto uvso:couleur ?couleur. 
	FILTER contains(?couleur, "noir").
	?auto uvso:nombreDePlaces ?places. 
	?places gr:aValeurEntier "5"^^xsd:int.
	?auto uvso:AFabricant ?marque. 
	FILTER (contains(str(?marque), "audi")).
	?auto uvso:StyleVehicule uvso:berline_occasion.
	?auto uvso:KiloetrageOdometre ?kilometrage.
	?kilometrage gr:aValeurFloat ?valeurKilometrage.
	FILTER (?valeurKilometrage <= 100000) .
	?auto uvo:Estimation ?estimation. 
	?estimation uvoo:aValeurMonetaire ?prix. 
	FILTER (?prix <= 100000 && ?prix >= 20000) .
 } LIMIT 10
		\end{lstlisting}
	\end{adjustbox}
	\vspace{-0.3cm}
	\caption{Une requête SPARQL en correspondance avec des préférences de l'utilisateur.}
	\label{fig_02}
	\vspace{-0.3cm}
\end{figure}

La mise en correspondance de motifs de graphes est essentiellement le mécanisme utilisé par SPARQL pour récupérer des informations à partir de graphes de connaissances RDF. Dans ce contexte, une contrainte est considérée comme une évaluation d'un motif de graphe sur le graphe de connaissances RDF. Pour trouver des solutions, les requêtes SPARQL peuvent utiliser des motifs de triplets et des modificateurs de solutions en tant que contraintes. Les motifs de triplets impliquent trois variables et les modificateurs de solutions tels que ORDER BY, DISTINCT et LIMIT peuvent être utilisés pour trier, éliminer les doublons et limiter des solutions. Cette approche bénéficie de l'expressivité des requêtes SPARQL, qui ont le pouvoir expressif de l'algèbre relationnelle.\\

\section{Expérimentations}\label{experiments}
\begin{figure}[htbp]
	\centering
	\begin{tabular}{c c}
		\includegraphics[width=0.20\textwidth]{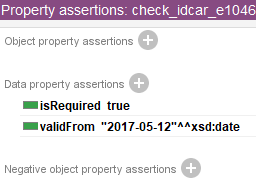} & \includegraphics[width=0.251\textwidth]{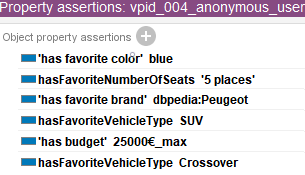}
	\end{tabular}
	
	\vspace{-0.2cm}
	\caption{Information déduite en utilisant les règles SWRL traduites à partir des contraintes  $\mathcal{C}_{KB1}$ et $\mathcal{C}_{KB2}$.}
	\label{fig_03}
	\vspace{-0.2cm}
\end{figure}

Afin d'évaluer l'approche proposée, nous utilisons le graphe de connaissances RDF composé de 5537 individus de descriptions de véhicules et de 367 préférences d'utilisateurs, qui contiennent un total de 822 000 triplets RDF basés sur les modèles de l'ontologie présentés dans \cite{le2022towards}. À partir d'une étude empirique des ensembles de données, nous montrons comment fonctionne notre SdR basé sur des contraintes via le graphe de connaissances RDF. Tout d'abord, nous organisons les préférences des utilisateurs et les descriptions de véhicules en triplets RDF basés sur le modèle de l'ontologie afin de collecter les données de manière formelle. La construction d'un SdR basé sur des contraintes se concentre ensuite sur la résolution de deux ensembles de contraintes : des contraintes de connaissances spécifiques au domaine et des contraintes de préférences utilisateur. En particulier, l'ensemble de contraintes basé sur les connaissances spécifiques au domaine est traduit en règles SWRL et directement implémenté sur le graphe de connaissances RDF via des modules de raisonnement. La déduction d'informations nouvelles et pertinentes sur chaque utilisateur et chaque élément de véhicule est ensuite ajoutée à l'ensemble de données comme illustré dans la figure \ref{fig_03}. L'ensemble de contraintes repose sur les préférences des utilisateurs liées aux informations sur leurs véhicules préférés, qui jouent un rôle essentiel dans la recherche de recommandations pertinentes. Par conséquent, nous formulons ces contraintes à l'aide de requêtes SPARQL basées sur la mise en correspondance de motifs sur des graphes et des modificateurs de solutions, comme indiqué dans la Figure \ref{fig_02}. Dans le cas idéal, toutes les variables peuvent être affectées et nous pouvons trouver des solutions vérifiant pour le graphe de connaissances RDF. Au final, les recommandations produites respectent l'ensemble des contraintes et sont donc les plus pertinentes.

\begin{figure}[htbp]
	\centerline{\includegraphics[width=0.45\textwidth]{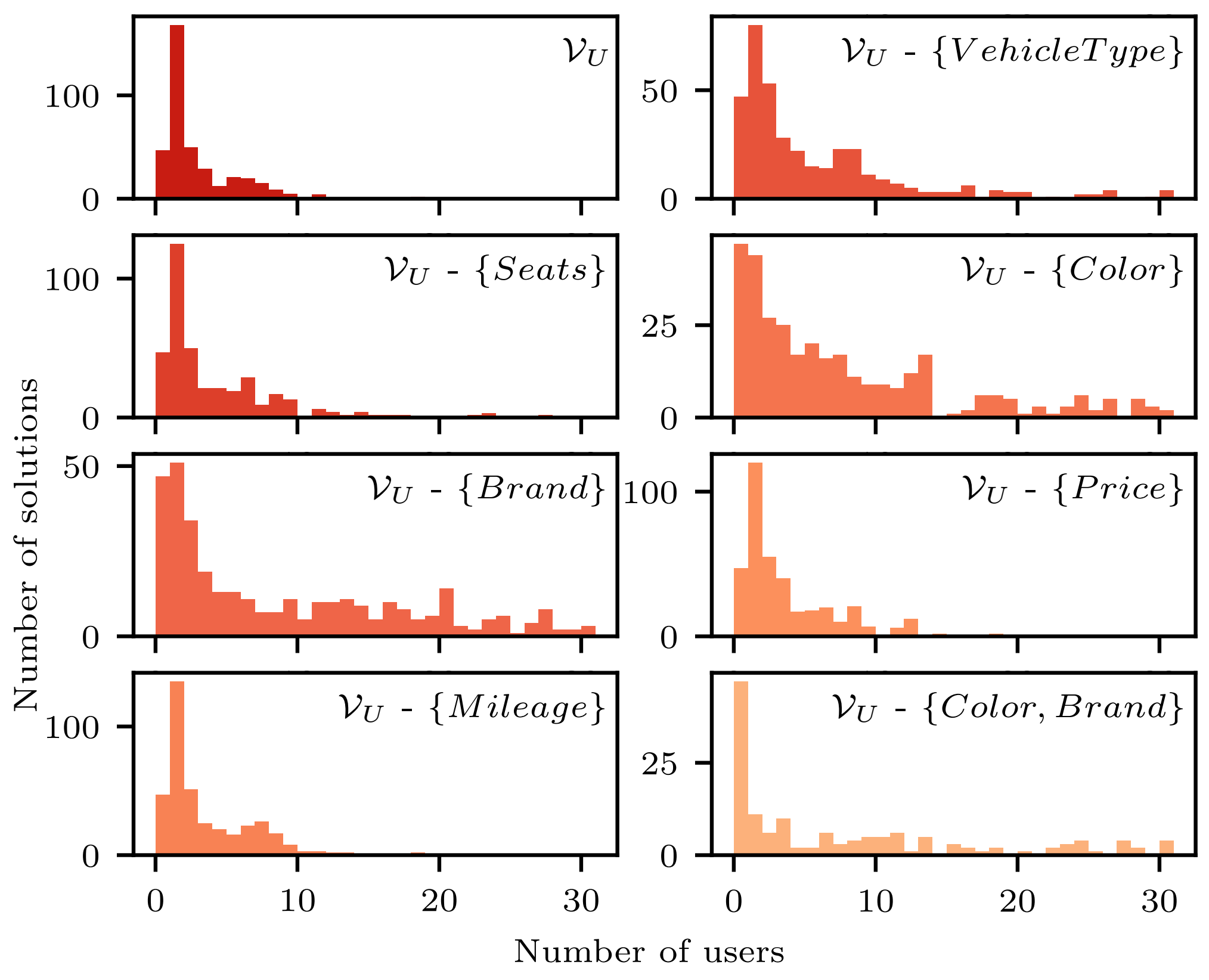}}
	\vspace{-0.5cm}
	\caption{Histogrammes représentant la distribution des solutions à travers différents ensembles de contraintes.}
	\label{fig_04}
	\vspace{-0.3cm}
\end{figure}

Cependant, de nombreux cas peuvent ne pas trouver de solution en raison de certaines incohérences entre les contraintes des préférences de l'utilisateur et les descriptions des véhicules. Il y a deux propositions possibles pour ce problème : (1) Enrichir et agrandir la base de connaissances RDF en augmentant le nombre de véhicules échangés sur les portails ; (2) Traiter et identifier un ensemble minimal de contraintes à partir des préférences de l'utilisateur. Au lieu d'enrichir le jeu de données comme dans la première proposition, la seconde proposition repose sur l'élimination ou l'adaptation des contraintes de l'utilisateur en utilisant un ensemble de diagnostic qui est défini comme un ensemble de contraintes $\Delta$ extrait de l'ensemble de contraintes $\mathcal{C}_F$ tel que les recommandations à partir du nouvel ensemble de contraintes $\mathcal{C}_F$$-\Delta$ sont cohérentes.

Afin d'expérimenter ce type de situation, nous avons construit des contraintes de préférences utilisateur basées sur l'ensemble de variables $\mathcal{V}_U$ $=$ $\{Seats,$ $VehicleType,$ $Brand,$ $Color,$ $Mileage,$ $Price\}$ extraites des préférences utilisateur dans l'ensemble de données. L'expérience avec toutes les contraintes basées sur les préférences de l'utilisateur a abouti à 88 \% des utilisateurs qui ont trouvé au moins une solution dans l'ensemble de données RDF. Avec la deuxième expérience, nous avons cherché à construire des ensembles de diagnostics afin de maximiser le nombre de solutions correspondant aux préférences de l'utilisateur et de réduire le nombre d'utilisateurs qui ne peuvent pas trouver de solution. Les ensembles de diagnostics incluent uniquement les contraintes éliminées de chaque préférence de l'utilisateur, en fonction d'un ordre de préférence défini par l'utilisateur pour ses préférences. Par exemple, $\Delta_1 = \{Places-Seats\}$, $\Delta_2 = \{TypeVehicule-VehicleType\}$, $\Delta_3 = \{Marque-Brand\}$, $\Delta_4 = \{Couleur-Color\}$, $\Delta_5 = \{Kilom\acute{e}trage-Mileage\}$, $\Delta_6 = \{Prix-Price\}$, $\Delta_7 = \{Couleur, Marque\}$. La figure \ref{fig_04} montre des histogrammes sur la distribution du nombre de solutions sur le nombre d'utilisateurs en utilisant différents ensembles de diagnostics. Avec toutes les contraintes $\mathcal{V}_U$, la majorité du nombre de solutions se situe dans une plage de 0 à 5 solutions par utilisateur. En appliquant des ensembles de diagnostics, le nombre de solutions s'étend avec une augmentation du nombre de solutions supérieure à 10 pour les utilisateurs. Ces changements dans le nombre de solutions sont particulièrement illustrés par l'ensemble de contraintes : $\mathcal{V}_U - \Delta_3$ et $\mathcal{V}_U - \Delta_4$. Pour réduire le nombre d'utilisateurs qui ne peuvent pas trouver de solution, l'élimination de plusieurs préférences de l'utilisateur peut devenir nécessaire. Cela signifie qu'il est nécessaire de faire un compromis entre la satisfaction de l'utilisateur et les résultats de recommandations.

Nous illustrons le SdR basé sur des contraintes à partir d'un graphe de connaissances RDF dans le domaine des véhicules. Les expérimentations menées confirment l'intérêt de notre approche de séparation des ensembles de contraintes en ensembles de contraintes de connaissances spécifiques au domaine et en ensembles de contraintes de préférences de l'utilisateur. En utilisant des règles SWRL, l'ensemble de connaissances spécifiques au domaine peut être déduit et intégré dans l'ensemble de données RDF. L'ensemble de contraintes construit sur les préférences utilisateur est traduit en requêtes SPARQL. Les recommandations pertinentes pour les utilisateurs sont extraites à partir des solutions obtenues par la recherche de motifs sur les graphes RDF.

\section{Conclusion et perspectives}\label{conclusion}
Dans cet article, nous avons présenté comment construire un SdR basé sur des contraintes s'appuyant sur un graphe de connaissances. Un tel système permet d'intégrer dans un modèle sémantique uniforme la description des éléments et celle du domaine dans lequel ils évoluent. Nous avons montré comment distinguer contraintes selon qu'elles concernent les connaissances spécifiques au domaine ou les préférences de l'utilisateur. En utilisant des règles SWRL, nous avons traduit les contraintes de connaissances spécifiques au domaine en règles et avons effectué des déductions de nouvelles informations pertinentes sur le graphe de connaissances RDF. Les contraintes de préférences de l'utilisateur peuvent être directement traduites en requêtes SPARQL. Nous avons mené une expérience sur notre approche basée sur le graphe de connaissances RDF de l'achat et de la vente de véhicules. Les résultats de recommandation obtenus à partir du SdR basé sur des contraintes sont prometteurs. Dans nos travaux futurs, nous prévoyons de rechercher l'exploitation d'ensembles de diagnostics, qui devraient être optimisés pour chaque utilisateur et pourraient aider à réduire le temps de performance pour proposer des recommandations pertinentes.
\section*{Remerciements}
Cette recherche a été financée par l’Agence National de la Recherche (ANR) et par l’entreprise Vivocaz au titre du projet France Relance – préservation de l’emploi R\&D (ANR-21-PRRD-0072-01).

\bibliographystyle{plain}
\bibliography{references} 

\begin{thebibliography}{10}

\bibitem{adomavicius2005toward}
Gediminas Adomavicius and Alexander Tuzhilin.
\newblock Toward the next generation of recommender systems: A survey of the
  state-of-the-art and possible extensions.
\newblock {\em IEEE transactions on knowledge and data engineering},
  17(6):734--749, 2005.

\bibitem{atas2019towards}
Muesluem Atas, TN~Trang Tran, Alexander Felfernig, Seda~Polat Erdeniz, Ralph
  Samer, and Martin Stettinger.
\newblock Towards similarity-aware constraint-based recommendation.
\newblock In {\em International Conference on Industrial, Engineering and Other
  Applications of Applied Intelligent Systems}, pages 287--299. Springer, 2019.

\bibitem{boudaa2021datatourist}
Boudjemaa Boudaa, Djamila Figuir, Slimane Hammoudi, and Sidi mohamed
  Benslimane.
\newblock Datatourist: A constraint-based recommender system using datatourisme
  ontology.
\newblock {\em International Journal of Decision Support System Technology},
  13(2):62--84, 2021.

\bibitem{bridge2005case}
Derek Bridge, Mehmet~H G{\"o}ker, Lorraine McGinty, and Barry Smyth.
\newblock Case-based recommender systems.
\newblock {\em The Knowledge Engineering Review}, 20(3), 2005.

\bibitem{dadouchi2022context}
Cam{\'e}lia Dadouchi, Bruno Agard, and Benoit Montreuil.
\newblock Context-aware interactive knowledge-based recommendation.
\newblock {\em SN Computer Science}, 3(6), 2022.

\bibitem{felfernig2016application}
Alexander Felfernig.
\newblock Application of constraint-based technologies in financial services
  recommendation.
\newblock In {\em CEUR Workshop}, 2016.

\bibitem{felfernig2008constraint}
Alexander Felfernig and Robin Burke.
\newblock Constraint-based recommender systems: technologies and research
  issues.
\newblock In {\em Proceedings of the 10th international conference on
  Electronic commerce}, 2008.

\bibitem{felfernig2006integrated}
Alexander Felfernig, Gerhard Friedrich, Dietmar Jannach, and Markus Zanker.
\newblock An integrated environment for the development of knowledge-based
  recommender applications.
\newblock {\em International Journal of Electronic Commerce}, 11(2):11--34,
  2006.

\bibitem{felfernig2011developing}
Alexander Felfernig, Gerhard Friedrich, Dietmar Jannach, and Markus Zanker.
\newblock Developing constraint-based recommenders.
\newblock In {\em Recommender systems handbook}, pages 187--215. Springer,
  2011.

\bibitem{felfernig2015constraint}
Alexander Felfernig, Gerhard Friedrich, Dietmar Jannach, and Markus Zanker.
\newblock Constraint-based recommender systems.
\newblock In {\em Recommender systems handbook}, pages 161--190. Springer,
  2015.

\bibitem{jannach2009constraint}
Dietmar Jannach, Markus Zanker, and Matthias Fuchs.
\newblock Constraint-based recommendation in tourism: A multiperspective case
  study.
\newblock {\em Information Technology \& Tourism}, 11(2):139--155, 2009.

\bibitem{le2022apport}
Luyen Le~Ngoc, Marie-H{\'e}l{\`e}ne Abel, and Philippe Gouspillou.
\newblock Apport des ontologies pour le calcul de la similarit{\'e}
  s{\'e}mantique au sein d'un syst{\`e}me de recommandation.
\newblock In {\em Ing{\'e}nierie des Connaissances (Ev{\`e}nement affili{\'e}
  {\`a} PFIA Plate-Forme Intelligence Artificielle)}, 2022.

\bibitem{le2022towards}
Luyen Le~Ngoc, Marie-H{\'e}l{\`e}ne Abel, and Philippe Gouspillou.
\newblock Towards an ontology-based recommender system for the vehicle sales
  area.
\newblock In {\em International Conference on Deep Learning, Artificial
  Intelligence and Robotics}, pages 126--136. Springer, 2022.

\bibitem{le2023constraint}
Luyen Le~Ngoc, Jinfeng Zhong, Elsa Negre, and Marie-H{\'e}l{\`e}ne Abel.
\newblock Constraint-based recommender system for crisis management
  simulations.
\newblock In {\em The 56th Hawaii International Conference on System Sciences},
  2023.

\bibitem{lu2015recommender}
Jie Lu, Dianshuang Wu, Mingsong Mao, Wei Wang, and Guangquan Zhang.
\newblock Recommender system application developments: a survey.
\newblock {\em Decision Support Systems}, 74:12--32, 2015.

\bibitem{mcguinness2004owl}
Deborah~L McGuinness, Frank Van~Harmelen, et~al.
\newblock Owl web ontology language overview.
\newblock {\em W3C recommendation}, 10(10):2004, 2004.

\bibitem{ramezani2008selecting}
Maryam Ramezani, Lawrence Bergman, Rich Thompson, Robin Burke, and Bamshad
  Mobasher.
\newblock Selecting and applying recommendation technology.
\newblock In {\em International Workshop on Recommendation and Collaboration in
  Conjunction with 2008 International ACM Conference on Intelligent User
  Interfaces, IUI}, pages 613--620, 2008.

\end{thebibliography}

\end{document}